\documentstyle[12pt]{article}
\begin{document}
\vspace*{-1in}
\begin{flushright}
CERN-TH. 7245/94 \\
DAMTP 94-31 \\
\end{flushright}
\vskip 70pt
\begin{center}
{\Large{\bf Chasing the light-gluino scenario through
 $b\rightarrow s \gamma$}}
\vskip 30pt
Gautam Bhattacharyya ${}^{*)}$
\vskip 10pt
{\it Theory Division, CERN, \\ CH 1211, Geneva 23,  SWITZERLAND}
\vskip 10pt
and
\vskip 10pt
Amitava Raychaudhuri
\vskip 10pt
{\it DAMTP, University of Cambridge,\\ Silver Street,
Cambridge CB3 9EW, U.K.\\and \\
Department of Pure Physics, University of Calcutta, \\
92 Acharya Prafulla Chandra Road, Calcutta 700 009, INDIA}
\vskip 70pt

{\bf ABSTRACT}
\end{center}

We investigate the impact of a light gluino,
which might have escaped detection at colliders,
on inclusive radiative $B$-decays mediated through
penguin-like diagrams.
We find that the
viability of the scenario depends largely on the magnitude of
the flavour-violating $c$-parameter. Some
previously allowed regions of parameter
space are now ruled out.
\vskip 25pt
\begin{center}
\end{center}
\vskip 40pt
\begin{flushleft}
CERN-TH. 7245/94 \\
DAMTP 94-31 \\
May 1994\\
\vskip 11pt
${}^{*)}$ gautam@cernvm.cern.ch
\end{flushleft}

\newpage
\setcounter{page} 1
 \pagestyle{plain}

There has been a continued speculation that a light gluino ($\sim$
2 -- 5 GeV) has escaped detection at the colliders [1].
This assertion had also been fueled by the observation that a light,
coloured, neutral fermion improves the agreement between low- and
high-energy $\alpha_S$ measurements; the light gluino is a strong
candidate to satisfy such a requirement. This possibility
has been looked into by a number of experiments [2], but it is
still very much open, crying out for verification. It is noteworthy
the direct search limits
on squark masses from the CDF collaboration at Fermilab
\cite{tevatron} are evaded in the presence of a light gluino; the
squarks need, in principle, to be heavier than only $M_Z/2$, from
non-observation at the CERN $e^+e^-$ collider, LEP.
However, it has been pointed out \cite{br2}
that the precision LEP measurements disfavour squarks below
60 GeV associated with such a light gluino. Of late, a particularly
interesting gateway to examine
various varieties of new physics, including this
speculative light-gluino scenario, has been provided
by the inclusive $B$-decay measurement, setting a limit
$Br(b\rightarrow s \gamma) < 5.4 \times 10^{-4}$ \cite{cleo}.
It has already
been pointed out [6,7] that this rare decay has a strong influence
on restricting the parameter space of supersymmetry (SUSY).
 This motivates
us to examine in this paper the present status of the light gluino
through this `microscope'. SUSY contributions to the rare decay
$b\rightarrow s \gamma$ have been examined in the
literature \cite{bbm} earlier.
On top of these investigations, we adopt a timely
specialization to the recently reheated issue of a light gluino,
following the improved experimental measurement.

The branching ratio of $b\rightarrow s\gamma$ is given
in units of the semileptonic $b$-decay branching ratio, as
\begin{equation}
 {{Br (b\rightarrow s\gamma)} \over{Br
(b\rightarrow ce\overline{\nu})}} = {{6 \alpha}\over{\pi\rho\lambda}}
\left|{{K_{tb} K_{ts}^*}\over{K_{bc}}}\right|^2
\left[\eta^{16/23} A_\gamma + {8\over3}(\eta^{14/23} - \eta^{16/23})
A_{g} + C\right]^2,
\label{brat}
\end{equation}
where $\eta = \alpha_S(M_Z)/\alpha_S(m_b) = 0.548$,
 $\rho = (1-8r^2+8r^6-r^8-24r^4 \mbox{\rm ln}r)$
 with $r = m_c/m_b$, $\lambda =
1 - 1.61\;\alpha_S(m_b)/\pi$, and $C (=-0.1766)$ is a coefficient from
a complete calculation of the leading-logarithmic QCD corrections
\cite{misiak}; $K$ is
the standard Cabibbo-Kobayashi-Maskawa matrix.
It may be noted that the ${m_b}^5$ dependence in
the partial decay widths of the $b$ quark cancels out in
eq. (\ref{brat}). An ${\cal{O}}(m^2_s/m^2_b)$ part in the
branching ratio is neglected.
We take $Br(b\rightarrow ce\overline{\nu}) = 0.107$.
$A_\gamma$ and $A_g$ are the coefficients of the effective operators
for $bs$-photon and $bs$-gluon interactions \cite{inamilim} following
from
\begin{equation}
{\cal L}_{eff} = {\sqrt{{G^2_F}\over{8\pi^3}}}
 K_{tb} K^*_{ts}~\overline{s}~
\sigma^{\mu\nu} \left[\sqrt{\alpha} A_\gamma F_{\mu\nu}
+ \sqrt{\alpha_S} A_g T_a G^a_{\mu\nu}\right] (m_b P_R
+ m_s P_L)~b.
\label{heff}
\end{equation}
The contributions to $A_\gamma$ and $A_g$ from $W$ bosons, charged
Higgs bosons and gauginos are listed in \cite{bg}.

The core of the interaction under our investigation is contained in a
particular subset of SUSY induced by the
quark-squark-gluino Lagrangian.
For the sake of making this note self-contained, we extract, in
what follows, the essence of the formalism of our earlier work
\cite{br2,br1}.  The quark-squark-gluino Lagrangian is given by
\begin{equation}
{\cal L}_{q{\widetilde q}{\widetilde g}} = i \sqrt{2} g_s {\widetilde
q}_i^{\dagger a} {\overline {\widetilde g}}_\alpha
(\lambda_\alpha/2)_{ab} \left[\Gamma_L^{ip} \frac{1-\gamma_5}{2}
 + \Gamma_R^{ip} \frac{1+\gamma_5}{2} \right] q_p^b,
\label{lag1}
\end{equation}
where, for three generations of quarks $p = 1 - 3$ , $i = 1 - 6$ (for
each quark flavour there are two squark states), the colour indices
$a,b = 1 - 3$ and $\alpha = 1 - 8$. The $(6\times 3)$ matrices
$\Gamma_L$ and $\Gamma_R$ are determined by the quark and squark mass
matrices shown below.

Flavour violation stems from the fact
that the quark and squark mass matrices are not
diagonal in the same basis. The  $(6\times 6)$
$\widetilde d$ mass squared matrix
(in a basis in which the $d$-quark mass matrix is diagonal) is
\begin{equation}
M_{\widetilde d}^2 = \pmatrix{~m_{0L}^2 I + \hat M_d^2
+c K \hat
 M_u^2 K^\dagger~ & ~ Am_{3/2}\hat M_d  \cr ~Am_{3/2}\hat M_d & ~
 m_{0R}^2 I +\hat M_d^2},
\label{matr}
\end{equation}
where $m_{0L}$ and $m_{0R}$ are flavour-blind supersymmetry-breaking
parameters for the left- and right-type squarks, respectively. (For
the sake of simplification, we have taken $m_{0L} = m_{0R} = m_0$ for
numerical purposes, which does not materially affect the conclusion
of the paper.) Here, $\hat M_u$ and $\hat
M_d$ are diagonal up- and down-quark mass matrices respectively. The
$c$-term corresponds to a quantum mass correction
for a $d$-type left squark driven by higgsino exchange.
 It may be noted that
$c$ is the most crucial parameter, originating from an electroweak
one-loop effect, which triggers  flavour-violating
interactions like $b\rightarrow s \gamma$.
In specific models $c$ can be estimated by the renormalization
group (RG) equations of the quark and squark mass parameters. In our
analysis $c$ is a phenomenological input.
The off-diagonal block in eq. (\ref{matr})
corresponds to left-right squark mixings
and is proportional to the $d$-type quark mass matrix.
 $\Gamma_L$ and $\Gamma_R$ in eq. (\ref{lag1}) are
\begin{equation}
\Gamma_L = \widetilde U
\pmatrix {~I \cr~0},~~\Gamma_R = \widetilde U \pmatrix
{~0 \cr~I };
\end{equation}
$\widetilde U$ is the matrix that diagonalises $M_{\widetilde d}^2$;
$m_{3/2}$ stands for the gravitino mass,
 and $I$  is the $(3\times 3)$
identity matrix. It should be mentioned that although
the above mass matrix is of the texture that follows from $N=1$
supergravity, a mild extension of the minimal
supersymmetric standard model (MSSM)
keeps the general structure unaltered.

When the $c$-induced SUSY interaction is turned on, $A_\gamma$
and $A_g$ in eq. (\ref{heff}) pick up terms in addition to those
given in \cite{bg}. Their modified expressions, denoted by
$A^\prime_\gamma$ and $A^\prime_g$, respectively, are given by:
\begin{eqnarray}
A^\prime_{\gamma} &=& A_\gamma +
 {{4}\over{9}} {{\alpha_S(M_Z)}\over{\alpha}}
\sin^2\theta_W M^2_W~S_\gamma,  \nonumber \\
A^\prime_g &=& A_g + {{\alpha_S(M_Z)}\over{6\alpha}}
\sin^2\theta_W M^2_W~S_g.
\label{ampl}
\end{eqnarray}
Although we compute with the complete
set of parameters, we present in the following
the expressions of $S_\gamma$ and $S_g$
in the simplified case when $A=0$:
\begin{equation}
S_{\gamma} = C_{11} + C_{21}
\label{amplph1}
\end{equation}
and
\begin{equation}
S_g = (C_{11} + C_{21})
+ 9 (\widetilde{C}_{11} + \widetilde{C}_{21})
\label{amplgl1}
\end{equation}
where the $C$- and $\widetilde{C}$-functions
are the three-point integrals \cite{pv}, the arguments
of which are the three external and the three internal masses of
the relevant penguins.
Generically, the $C$-functions correspond to the case when a
photon (or a gluon) couples to the internal squark lines in the
penguin diagrams, while the $\widetilde{C}$-functions refer to
the situation when a gluon is emitted from an internal gluino line.
The $C$- and $\widetilde{C}$-functions in eqs. (\ref{amplph1}) and
(\ref{amplgl1}) represent their final forms after the super-GIM
subtraction (generically,
$C \equiv C(m^2_{\widetilde{b}})-C(m^2_{\widetilde{d}})$ and
$\widetilde{C} \equiv
\widetilde{C}(m^2_{\widetilde{b}})-\widetilde{C}(m^2_{\widetilde{d}})$).
Both $C$ and $\widetilde{C}$
are proportional to $cm^2_t$, the mass splitting between
$\widetilde{b}_L$ and any of the remaining $d$-type squarks,
controlling the rate of flavour violation. (In the actual
calculation, the GIM-subtraction is done numerically.)
 To evaluate the three-point functions
we use the code developed in \cite{biswarup} and employed subsequently
in \cite{br2,br1}. We also cross-check our calculation by performing
a systematic expansion in powers of the ratios of the masses of the
light and heavy particles. The approximate expressions of $S_\gamma$
and $S_g$ used in eq. (\ref{ampl}), which agree within $1\%$
with those in eqs. (\ref{amplph1}) and (\ref{amplgl1}),
are shown below ($ x = m^2_{\widetilde{g}}/m^2_0 $,
where $m_{\widetilde{g}}$ is the mass of the gluino):
\begin{equation}
 S_\gamma = {{c m^2_t}\over{6 m^4_0}} \left[(x-1)^{-4} (1-8x-17x^2)
 + 6(x-1)^{-5} x^2 (x+3) \ln x\right]
\label{amplph2}
\end{equation}
and
\begin{eqnarray}
 S_g &= &{{c m^2_t}\over{6 m^4_0}} \left[(x-1)^{-4} (x^2+172x+19)
 + 6x(x-1)^{-5}(x^2-15x-18) \ln x\right].
\label{amplgl2}
\end{eqnarray}

The results of our analysis are presented in fig. 1. To appreciate the
effect of the light gluino in
the context of the full theory of SUSY, we have included the
contributions of the charged Higgs and the gauginos. For this, we
have assumed the same simplified limit $\mu = 0$ and $\tan\beta = 1$
as in \cite{bg}, which is in agreement with the light-gluino
scenario.
Results are presented for three different values of the parameter $c$.
The broken line corresponds to choosing $c=0$, i.e. no contribution
from the gluino sector at all.
The SM contribution for $m_t = 180$ GeV is also shown as the dotted line
(the $m_t$-dependence of the branching ratio is rather mild).
$m_{\widetilde{g}}$ is set to 3 GeV in our analysis.

Since $c<0$ is preferred
in the MSSM, the lightest of the $\widetilde{d}$-type squarks,
dominantly $\widetilde{b}_L$, has a mass $\simeq \sqrt{m_0^2 + c m_t^2}$
(for $A = 0$).
Thus for a given choice of  $m_0$
and for a fixed $m_t$, the maximum magnitude of $c$ is
restricted by the LEP bound $\sqrt{m_0^2 + c m_t^2} \geq~45$ GeV.
For $m_t = 180$ GeV and $m_0 = 60$ GeV, this requires
$|c| \leq 0.05$.
Now, choosing $c = -0.05$ and
a charged Higgs mass ($M_{H^+}$) equal to 100 GeV, it becomes
evident from fig. 1 that the squark-gluino contribution
dominates over the rest for $m_0 < 100$ GeV.
The figure corresponds to the situation when there is no left-right
squark mixing, i.e. $A = 0$.
Under these circumstances, if one uses the CLEO
bound $Br(b\rightarrow s \gamma) < 5.4 \times 10^{-4}$, the light
gluino is completely disfavoured, no matter what the squark mass
is. It ought to be stressed that such a choice of $c$ is in
good consonance with the predictions from the RG evolution of the squark
masses \cite{hagelin}.
If one chooses $c=-0.01$, then for the same $m_t$ and
$M_{H^+}$, the gluino-induced effect suffers a significant reduction,
ensuring the viability of a light gluino.
Moreover, a larger value of $M_{H^+}$,
say 500 GeV, reduces the $H^+$ contribution to a large extent, leaving
ample room for a light gluino to be accommodated for $m_0>65$ GeV,
even with a choice of $c = -0.05$ and $m_t = 180$ GeV. Other parameters
remaining the same, choosing $A = 3$ decreases the effect very slightly,
at most by $\sim 2 \%$. If one deviates
from the MSSM and assumes a positive value for $c$,
the gluino-induced effect
becomes less prominent as a result of its destructive interference
with the other sectors, and no significant bound
could be set at all. It may be noted that varying $m_{\widetilde{g}}$
in the range (1 -- 5) GeV has no numerical impact within
the scale of the figure.

We conclude that the inclusive $b\rightarrow s \gamma$ measurement
imposes a stringent constraint on the light-gluino scenario for
a reasonable choice of the model parameters.
 For example, for $c = -0.05$,
$m_t = 180$ GeV and $M_{H^+} = 100$ GeV, the light-gluino window
is virtually closed for arbitrary choices of the squark masses.
Needless to say,
the sign and the magnitude of $c$, for which there is a significant
freedom, has a crucial role to play in drawing such a conclusion.
On the other hand, the consequence of a
light gluino in the MSSM, in the
context of unification of gauge and Yukawa couplings, has been shown
\cite{marcela} to pose a very tight restriction on the allowed
values of $\alpha_S(M_Z)$,
keeping it consistent, nevertheless, with the
prediction at LEP. Additionally, if one demands the breakdown
of electroweak symmetry radiatively in the MSSM (irrespective of
the criterion of unification), a light gluino is difficult to be
accommodated \cite{lopez}.
This analysis, which probes a rather direct contribution
of a light gluino, concludes that the window is
still open, albeit with a smaller region of allowed parameter space.
Further investigation and more accurate experimental
measurements are therefore called for before any final verdict
can be drawn on this issue.

\vskip 10pt
\noindent{\bf Acknowledgements}
\par

GB gratefully acknowledges interesting and stimulating discussions with
M. Carena and C. Wagner. Both authors thank D.P. Roy
for making some critical remarks. The work of AR
is supported in part by C.S.I.R. (India), D.S.T. (India) and
the Commission of the European Communities. He is grateful to
J.C. Taylor and the DAMTP, University of Cambridge, for hospitality.

\newpage
\noindent{\bf Figure caption}
\par

\vskip 50pt
\begin{itemize}
\item [1.]
{}~~The branching ratio
 for the process $b \rightarrow s \gamma$ as a function
of the average squark mass ($m_0$) for different values of the
flavour-violation
parameter $c$ (solid lines).  Also shown are the branching ratio
with no contribution from the gluino sector (broken line) and
from the standard model alone (dotted line).
\end{itemize}


\newpage

\begin{thebibliography}{99}

\bibitem{cla1} L. Clavelli, F.W. Coulter and K.Yuan, {\it Phys. Rev.}
{\bf D47}, 1973 (1993); \\
 L. Clavelli et al., {\it Phys. Lett.} {\bf B291},
426 (1992);\\ M. Je\.{z}abek and J.H. K\"{u}hn, {\it ibid.}
{\bf B301}, 121 (1993); \\ R.G. Roberts and W.J. Stirling,
{\it ibid.} {\bf B313}, 453 (1993);
\\ J. Ellis, D.V. Nanopoulos and D.A. Ross,
{\it ibid.} {\bf B305}, 375 (1993).

\bibitem{e-2} UA1 Collaboration, C. Albajar et al., {\it Phys.
Lett.} {\bf B198}, 261 (1987) and references therein; \\
HELIOS Collaboration, T. Akesson et al.,
{\it Z. Phys.} {\bf C52}, 219 (1991); \\
NA3 Collaboration, J.P. Dishaw et al., {\it Phys.
Lett.} {\bf B85}, 142 (1979).

\bibitem{tevatron} CDF Collaboration, F. Abe et al.,
{\it Phys. Rev. Lett.} {\bf 69}, 3439 (1992).

\bibitem{br2} G. Bhattacharyya and A. Raychaudhuri, {\it Phys. Rev.}
{\bf D49}, R1156 (1994).

\bibitem{cleo} CLEO Collaboration, R. Ammar et al.,
{\it Phys. Rev. Lett.} {\bf 71}, 674 (1993); \\
T. Browder, K. Honscheid
and  S. Playfer, in ``{\it B Decays}", 2nd edition, ed. S. Stone
(World Scientific, Singapore, 1994).

\bibitem{barger} V. Barger, M.S. Berger and R.J.N. Phillips,
{\it Phys. Rev. Lett.} {\bf 70}, 1368 (1993); \\
J.L. Hewett, {\it ibid.} {\bf 70}, 1045 (1993).

\bibitem{bg} R. Barbieri and G.F. Giudice, {\it Phys. Lett.}
{\bf B309}, 86 (1993).

\bibitem{bbm} S. Bertolini, F. Borzumati and A. Masiero,
{\it Phys. Lett.} {\bf B192}, 437 (1987); \\ S. Bertolini,
F. Borzumati, A. Masiero and G. Ridolfi, {\it Nucl. Phys.}
{\bf B353}, 591 (1991); \\ B. Mukhopadhyaya
and S. Raychaudhuri, {\it Z. Phys.} {\bf C45}, 421 (1990); \\
F.M. Borzumati, DESY report DESY 93-090 (unpublished); \\
S. Bertolini and F. Vissani, SISSA report SISSA 40/94/EP (unpublished).

\bibitem{misiak} M. Misiak, {\it Phys. Lett.} {\bf B269}, 161 (1991).

\bibitem{inamilim} T. Inami and C.S. Lim, {\it Prog. Theor. Phys.}
{\bf 65}, 297 (1981).

\bibitem{br1} G. Bhattacharyya and A. Raychaudhuri, {\it Phys. Rev.}
{\bf D47}, 2014 (1993).

\bibitem{pv} G. 't Hooft and M. Veltman, {\it Nucl.
Phys.} {\bf B153}, 365 (1979); \\ G. Passarino and M. Veltman, {\it
ibid.} {\bf B160}, 151 (1979).

\bibitem{biswarup} B. Mukhopadhyaya and A. Raychaudhuri,
{\it Phys. Rev.} {\bf D39}, 280 (1989); \\
 A. Raychaudhuri, work in progress.

\bibitem{hagelin} J.S. Hagelin, S. Kelley and T. Tanaka, MIU report
MIU-THP-92/59 (unpublished). We thank G. Giudice for bringing
this paper to our notice.

\bibitem{marcela} M. Carena et al., {\it Phys. Lett.} {\bf B317},
346 (1993).

\bibitem{lopez} J.L. Lopez, D.V. Nanopoulos and X. Wang,
{\it Phys. Lett.} {\bf B313}, 241 (1993); \\
M.A. D\'{i}az, Vanderbilt report VAND-TH-94-7 (unpublished).


\end{thebibliography}
\end{document}